\documentstyle[prd,floats,aps,epsf,twocolumn]{revtex}
\input epsf

\newcommand {\abs}[1]{\mid \!\! #1 \!\! \mid}
\def \cal {\mathcal}
\def \met {{\,/\!\!\!\!E_{T}}}

\def \begineq {\begin{equation}}
\def \endeq {\end{equation}}

\def \ltapprox {\,\raisebox{-0.6ex}{$\stackrel{<}{\sim}$}\,}

\def \Sleuth   {{Sleuth}}

\def \jj {\,2j}

\def \jjjj {\,4j}

\newcommand{\guess}[1]
{\underline{#1}}

\def \wwemubkg {\unknown}
\def \wwemubkgerror {\unknown}

\def \wwemusigeff {\unknown}
\def \wwemudata {\unknown}
\def \wwemuxseclimit {\unknown~pb}

\def \zzcbkg {\unknown}
\def \zzcbkgerror {\unknown}

\def \zzcsigeff {\unknown}
\def \zzcdata {\unknown}
\def \zzcxseclimit {\unknown~pb}

\def \hwwcabkg {\unknown}
\def \hwwcabkgerror {\unknown}

\def \hwwcasigeff {\unknown}
\def \hwwcadata {\unknown}
\def \hwwcaxseclimit {\unknown~pb}
\def \hwwcbbkg {\unknown}
\def \hwwcbbkgerror {\unknown}

\def \hwwcbsigeff {\unknown}
\def \hwwcbdata {\unknown}
\def \hwwcbxseclimit {\unknown~pb}
\def \hwwccbkg {\unknown}
\def \hwwccbkgerror {\unknown}

\def \hwwccsigeff {\unknown}
\def \hwwccdata {\unknown}
\def \hwwccxseclimit {\unknown~pb}

\def \hzzcabkg {\unknown}
\def \hzzcabkgerror {\unknown}

\def \hzzcasigeff {\unknown}
\def \hzzcadata {\unknown}
\def \hzzcaxseclimit {\unknown~pb}
\def \hzzcbbkg {\unknown}
\def \hzzcbbkgerror {\unknown}

\def \hzzcbsigeff {\unknown}
\def \hzzcbdata {\unknown}
\def \hzzcbxseclimit {\unknown~pb}
\def \hzzccbkg {\unknown}
\def \hzzccbkgerror {\unknown}

\def \hzzccsigeff {\unknown}
\def \hzzccdata {\unknown}
\def \hzzccxseclimit {\unknown~pb}

\def \whbkg {\unknown}
\def \whbkgerror {\unknown}

\def \whsigeff {\unknown}
\def \whdata {\unknown}
\def \whxseclimit {\unknown~pb}

\def \zhbkg {\unknown}
\def \zhbkgerror {\unknown}

\def \zhsigeff {\unknown}
\def \zhdata {\unknown}
\def \zhxseclimit {\unknown~pb}

\def \ttemetbkg {\unknown}
\def \ttemetbkgerror {\unknown}

\def \ttemetsigeff {\unknown}
\def \ttemetdata {\unknown}
\def \ttemetxseclimit {\unknown~pb}
\def \ttemetxsec {\unknown~pb}
\def \ttemetxsecLowerError {\unknown~pb}
\def \ttemetxsecUpperError {\unknown~pb}

\def \ttemubkg {\unknown}
\def \ttemubkgerror {\unknown}

\def \ttemusigeff {\unknown}
\def \ttemudata {\unknown}
\def \ttemuxseclimit {\unknown~pb}
\def \ttemuxsec {\unknown~pb}
\def \ttemuxsecLowerError {\unknown~pb}
\def \ttemuxsecUpperError {\unknown~pb}

\def \zttabkg {\unknown}
\def \zttabkgerror {\unknown}

\def \zttasigeff {\unknown}
\def \zttadata {\unknown}
\def \zttaxseclimit {\unknown~pb}
\def \zttebkg {\unknown}
\def \zttebkgerror {\unknown}

\def \zttesigeff {\unknown}
\def \zttedata {\unknown}
\def \zttexseclimit {\unknown~pb}

\def \lqcbkg {\unknown}
\def \lqcbkgerror {\unknown}

\def \lqcsigeff {\unknown}
\def \lqcdata {\unknown}
\def \lqcxseclimit {\unknown~pb}

\def \wprimeabkg {\unknown}
\def \wprimeabkgerror {\unknown}

\def \wprimeasigeff {\unknown}
\def \wprimeadata {\unknown}
\def \wprimeaxseclimit {\unknown~pb}
\def \wprimedbkg {\unknown}
\def \wprimedbkgerror {\unknown}

\def \wprimedsigeff {\unknown}
\def \wprimeddata {\unknown}
\def \wprimedxseclimit {\unknown~pb}
\def \wprimegbkg {\unknown}
\def \wprimegbkgerror {\unknown}

\def \wprimegsigeff {\unknown}
\def \wprimegdata {\unknown}
\def \wprimegxseclimit {\unknown~pb}

\def \wwemubkg {19.0}
\def \wwemubkgerror {\phantom04.0}

\def \wwemusigeff {0.14}
\def \wwemudata {23}
\def \wwemuxseclimit {1.1\phantom0~pb}

\def \zzcbkg {19.7}
\def \zzcbkgerror {\phantom04.1}

\def \zzcsigeff {0.12}
\def \zzcdata {19}
\def \zzcxseclimit {0.8\phantom0~pb}

\def \ttemetbkg {\phantom03.1}
\def \ttemetbkgerror {\phantom00.9}

\def \ttemetsigeff {0.13}
\def \ttemetdata {8}
\def \ttemetxseclimit {0.8\phantom0~pb}
\def \ttemetxsec {0.39}
\def \ttemetxsecLowerError {0.19}
\def \ttemetxsecUpperError {0.21}

\def \ttemubkg {\phantom00.6}
\def \ttemubkgerror {\phantom00.2}

\def \ttemusigeff {0.14}
\def \ttemudata {2}
\def \ttemuxseclimit {0.4\phantom0~pb}
\def \ttemuxsec {0.14}
\def \ttemuxsecLowerError {0.08}
\def \ttemuxsecUpperError {0.15}

\def \hwwcabkg {29.6}
\def \hwwcabkgerror {\phantom06.5}

\def \hwwcasigeff {0.02}
\def \hwwcadata {32}
\def \hwwcaxseclimit {11.0\phantom0~pb}

\def \hwwcbbkg {66.0}
\def \hwwcbbkgerror {13.8}

\def \hwwcbsigeff {0.07}
\def \hwwcbdata {69}
\def \hwwcbxseclimit {4.4\phantom0~pb}

\def \hwwccbkg {43.1}
\def \hwwccbkgerror {\phantom09.2}

\def \hwwccsigeff {0.06}
\def \hwwccdata {44}
\def \hwwccxseclimit {3.6\phantom0~pb}

\def \hzzcabkg {17.9}
\def \hzzcabkgerror {\phantom03.7}

\def \hzzcasigeff {0.15}
\def \hzzcadata {15}
\def \hzzcaxseclimit {0.6\phantom0~pb}

\def \hzzcbbkg {18.8}
\def \hzzcbbkgerror {\phantom03.8}

\def \hzzcbsigeff {0.15}
\def \hzzcbdata {12}
\def \hzzcbxseclimit {0.4\phantom0~pb}

\def \hzzccbkg {18.1}
\def \hzzccbkgerror {\phantom03.7}

\def \hzzccsigeff {0.17}
\def \hzzccdata {18}
\def \hzzccxseclimit {0.6\phantom0~pb}

\def \wprimeabkg {27.7}
\def \wprimeabkgerror {\phantom06.3}

\def \wprimeasigeff {0.05}
\def \wprimeadata {29}
\def \wprimeaxseclimit {3.4\phantom0~pb}

\def \wprimedbkg {22.7}
\def \wprimedbkgerror {\phantom05.2}

\def \wprimedsigeff {0.23}
\def \wprimeddata {27}
\def \wprimedxseclimit {0.7\phantom0~pb}

\def \wprimegbkg {\phantom02.1}
\def \wprimegbkgerror {\phantom00.8}

\def \wprimegsigeff {0.26}
\def \wprimegdata {2}
\def \wprimegxseclimit {0.2\phantom0~pb}

\def \zttabkg {18.7}
\def \zttabkgerror {\phantom04.0}

\def \zttasigeff {0.11}
\def \zttadata {20}
\def \zttaxseclimit {1.1\phantom0~pb}

\def \zttebkg {18.7}
\def \zttebkgerror {\phantom04.0}

\def \zttesigeff {0.14}
\def \zttedata {20}
\def \zttexseclimit {0.9\phantom0~pb}

\def \zttgbkg {\phantom03.8}
\def \zttgbkgerror {\phantom01.0}

\def \zttgsigeff {0.14}
\def \zttgdata {2}
\def \zttgxseclimit {0.3\phantom0~pb}

\def \whbkg {37.3}
\def \whbkgerror {\phantom08.2}

\def \whsigeff {0.08}
\def \whdata {32}
\def \whxseclimit {2.0\phantom0~pb}

\def \zhbkg {19.5}
\def \zhbkgerror {\phantom04.1}

\def \zhsigeff {0.20}
\def \zhdata {25}
\def \zhxseclimit {0.8\phantom0~pb}

\def \lqcbkg {\phantom00.3}
\def \lqcbkgerror {\phantom00.1}

\def \lqcsigeff {0.33}
\def \lqcdata {0}
\def \lqcxseclimit {0.07~pb}

\def \Quaero {{\sc quaero}}
\def \Sleuth {{\sc sleuth}}

\begin{document}

\onecolumn

\title{Search for New Physics Using {\bf \Quaero}: A General Interface to D\O\ Event Data}

%
\author{                                                                      
V.M.~Abazov,$^{23}$                                                           
B.~Abbott,$^{58}$                                                             
A.~Abdesselam,$^{11}$                                                         
M.~Abolins,$^{51}$                                                            
V.~Abramov,$^{26}$                                                            
B.S.~Acharya,$^{17}$                                                          
D.L.~Adams,$^{60}$                                                            
M.~Adams,$^{38}$                                                              
S.N.~Ahmed,$^{21}$                                                            
G.D.~Alexeev,$^{23}$
% ...
\\                                                                            
\vskip 0.30cm                                                                 
\centerline{(D\O\ Collaboration)}                                             
\vskip 0.30cm                                                                 
}                                                                             
\address{                                                                     
\centerline{$^{1}$Universidad de Buenos Aires, Buenos Aires, Argentina}       
\centerline{$^{2}$LAFEX, Centro Brasileiro de Pesquisas F{\'\i}sicas,         
                  Rio de Janeiro, Brazil}                                     
\centerline{$^{3}$Universidade do Estado do Rio de Janeiro,                   
                  Rio de Janeiro, Brazil}                                     
\centerline{$^{4}$Institute of High Energy Physics, Beijing,                  
                  People's Republic of China}                                 
\centerline{$^{5}$Universidad de los Andes, Bogot\'{a}, Colombia}             
\centerline{$^{6}$Charles University, Center for Particle Physics,            
                  Prague, Czech Republic}                                     
\centerline{$^{7}$Institute of Physics, Academy of Sciences, Center           
                  for Particle Physics, Prague, Czech Republic}               
\centerline{$^{8}$Universidad San Francisco de Quito, Quito, Ecuador}         
\centerline{$^{9}$Institut des Sciences Nucl\'eaires, IN2P3-CNRS,             
                  Universite de Grenoble 1, Grenoble, France}                 
\centerline{$^{10}$CPPM, IN2P3-CNRS, Universit\'e de la M\'editerran\'ee,     
                  Marseille, France}                                          
\centerline{$^{11}$Laboratoire de l'Acc\'el\'erateur Lin\'eaire,              
                  IN2P3-CNRS, Orsay, France}                                  
\centerline{$^{12}$LPNHE, Universit\'es Paris VI and VII, IN2P3-CNRS,         
                  Paris, France}                                              
\centerline{$^{13}$DAPNIA/Service de Physique des Particules, CEA, Saclay,    
                  France}                                                     
\centerline{$^{14}$Universit{\"a}t Mainz, Institut f{\"u}r Physik,            
                  Mainz, Germany}                                             
\centerline{$^{15}$Panjab University, Chandigarh, India}                      
\centerline{$^{16}$Delhi University, Delhi, India}                            
\centerline{$^{17}$Tata Institute of Fundamental Research, Mumbai, India}     
\centerline{$^{18}$Seoul National University, Seoul, Korea}                   
\centerline{$^{19}$CINVESTAV, Mexico City, Mexico}                            
\centerline{$^{20}$FOM-Institute NIKHEF and University of                     
                  Amsterdam/NIKHEF, Amsterdam, The Netherlands}               
\centerline{$^{21}$University of Nijmegen/NIKHEF, Nijmegen, The               
                  Netherlands}                                                
\centerline{$^{22}$Institute of Nuclear Physics, Krak\'ow, Poland}            
\centerline{$^{23}$Joint Institute for Nuclear Research, Dubna, Russia}       
\centerline{$^{24}$Institute for Theoretical and Experimental Physics,        
                   Moscow, Russia}                                            
\centerline{$^{25}$Moscow State University, Moscow, Russia}                   
\centerline{$^{26}$Institute for High Energy Physics, Protvino, Russia}       
\centerline{$^{27}$Lancaster University, Lancaster, United Kingdom}           
\centerline{$^{28}$Imperial College, London, United Kingdom}                  
\centerline{$^{29}$University of Arizona, Tucson, Arizona 85721}              
\centerline{$^{30}$Lawrence Berkeley National Laboratory and University of    
                  California, Berkeley, California 94720}                     
\centerline{$^{31}$University of California, Davis, California 95616}         
\centerline{$^{32}$California State University, Fresno, California 93740}     
\centerline{$^{33}$University of California, Irvine, California 92697}        
\centerline{$^{34}$University of California, Riverside, California 92521}     
\centerline{$^{35}$Florida State University, Tallahassee, Florida 32306}      
\centerline{$^{36}$University of Hawaii, Honolulu, Hawaii 96822}              
\centerline{$^{37}$Fermi National Accelerator Laboratory, Batavia,            
                   Illinois 60510}                                            
\centerline{$^{38}$University of Illinois at Chicago, Chicago,                
                   Illinois 60607}                                            
\centerline{$^{39}$Northern Illinois University, DeKalb, Illinois 60115}      
\centerline{$^{40}$Northwestern University, Evanston, Illinois 60208}         
\centerline{$^{41}$Indiana University, Bloomington, Indiana 47405}            
\centerline{$^{42}$University of Notre Dame, Notre Dame, Indiana 46556}       
\centerline{$^{43}$Iowa State University, Ames, Iowa 50011}                   
\centerline{$^{44}$University of Kansas, Lawrence, Kansas 66045}              
\centerline{$^{45}$Kansas State University, Manhattan, Kansas 66506}          
\centerline{$^{46}$Louisiana Tech University, Ruston, Louisiana 71272}        
\centerline{$^{47}$University of Maryland, College Park, Maryland 20742}      
\centerline{$^{48}$Boston University, Boston, Massachusetts 02215}            
\centerline{$^{49}$Northeastern University, Boston, Massachusetts 02115}      
\centerline{$^{50}$University of Michigan, Ann Arbor, Michigan 48109}         
\centerline{$^{51}$Michigan State University, East Lansing, Michigan 48824}   
\centerline{$^{52}$University of Nebraska, Lincoln, Nebraska 68588}           
\centerline{$^{53}$Columbia University, New York, New York 10027}             
\centerline{$^{54}$University of Rochester, Rochester, New York 14627}        
\centerline{$^{55}$State University of New York, Stony Brook,                 
                   New York 11794}                                            
\centerline{$^{56}$Brookhaven National Laboratory, Upton, New York 11973}     
\centerline{$^{57}$Langston University, Langston, Oklahoma 73050}             
\centerline{$^{58}$University of Oklahoma, Norman, Oklahoma 73019}            
\centerline{$^{59}$Brown University, Providence, Rhode Island 02912}          
\centerline{$^{60}$University of Texas, Arlington, Texas 76019}               
\centerline{$^{61}$Texas A\&M University, College Station, Texas 77843}       
\centerline{$^{62}$Rice University, Houston, Texas 77005}                     
\centerline{$^{63}$University of Virginia, Charlottesville, Virginia 22901}   
\centerline{$^{64}$University of Washington, Seattle, Washington 98195}       
}                                                                             

\maketitle
\vskip 10pt

{\samepage
{\bf
\begin{center}
Abstract
\end{center}
}
\begin{center}
\begin{minipage}{.8\textwidth}
{\small 

We describe \Quaero, a method that i) enables the automatic optimization of searches for physics beyond the standard model, and ii) provides a mechanism for making high energy collider data generally available.  We apply \Quaero\ to searches for standard model $WW$, $ZZ$, and $t\bar{t}$ production, and to searches for these objects produced through a new heavy resonance.  Through this interface, we make three data sets collected by the D\O\ experiment at $\sqrt{s}=1.8$~TeV publicly available.

}
\end{minipage}
\end{center}
}


\twocolumn

\renewcommand {\section}[1] {}
\renewcommand {\subsection}[1] {}
\renewcommand {\subsubsection}[1] {}

\section{Introduction and Motivation}

It is generally recognized that the standard model, a successful description of the fundamental particles and their interactions, must be incomplete.  Models that extend the standard model often predict rich phenomenology at the scale of a few hundred GeV, an energy regime accessible to the Fermilab Tevatron.  Due in part to the complexity of the apparatus required to test models at such large energies, experimental responses to these ideas have not kept pace.  Any technique that reduces the time required to test a particular candidate theory would allow more such theories to be tested, reducing the possibility that the data contain overlooked evidence for new physics.

Once data are collected and the backgrounds have been understood, the testing of any specific model in principle follows a well-defined procedure.  In practice, this process has been far from automatic.  Even when the basic selection criteria and background estimates are taken from a previous analysis, the reinterpretation of the data in the context of a new model often requires a substantial length of time.

Ideally, the data should be ``published'' in such a way that others in the community can easily use those data to test a variety of models.  The publishing of experimental distributions in journals allows this to occur at some level, but an effective publishing of a multidimensional data set has, to our knowledge, not yet been accomplished by a large particle physics experiment.  The problem appears to be that such data are context-specific, requiring detailed knowledge of the complexities of the apparatus.  This knowledge must somehow be incorporated either into the data or into whatever tool the non-expert would use to analyze those data. 

Many data samples and backgrounds have been defined in the context of \Sleuth~\cite{SherlockPRD1}, a quasi-model-independent search strategy for new high $p_T$ physics that has been applied to a number of exclusive final states~\cite{SherlockPRL,SherlockPRD2} in the data collected by the D\O\ detector~\cite{D0Detector} during 1992--1996 in Run I of the Fermilab Tevatron.  In this Letter we describe a tool (\Quaero) that automatically optimizes an analysis for a particular signature, using these samples and standard model backgrounds.  \Sleuth\ and \Quaero\ are complementary approaches to searches for new phenomena, enabling analyses that are both general (\Sleuth) and focused (\Quaero).   We demonstrate the use of \Quaero\ in eleven separate searches: standard model $WW$ and $ZZ$ production; standard model $t\bar{t}$ production with leptonic and semileptonic decays; resonant $WW$, $ZZ$, $WZ$, and $t\bar{t}$ production; associated Higgs boson production; and pair production of first generation scalar leptoquarks.  The data described here are accessible through \Quaero\ on the World Wide Web~\cite{Quaero}, for general use by the particle physics community.

\section{Algorithm}
The signals predicted by most theories of physics beyond the standard model involve an increased number of predicted events in some region of an appropriate variable space.  In this case the optimization of the analysis can be understood as the selection of the region in this variable space that minimizes $\overline{\sigma^{95\%}}$, the expected 95\% confidence level (CL) upper limit on the cross section of the signal in question, assuming the data contain no signal.  The optimization algorithm consists of a few simple steps:
\newcounter{romanlistc}
\setcounter{romanlistc}{0}
\begin{list}{$($\roman{romanlistc}$)$}
{\usecounter{romanlistc}
\setlength{\parsep}{0pt}
\setlength{\itemsep}{0pt}}
\item Kernel density estimation~\cite{Scott} is used to estimate the probability distributions $p(\vec{x}|s)$ and $p(\vec{x}|b)$ for the signal and background samples in a low-dimensional variable space $\cal V$, where $\vec{x}\in{\cal V}$. The signal sample is contained in a Monte Carlo file provided as input to \Quaero.  The background sample is constructed from all known standard model and instrumental sources.
\item A discriminant function $D(\vec{x})$ is defined by~\cite{PDE}
\begin{equation}
D(\vec{x}) = \frac{p(\vec{x}|s)}{p(\vec{x}|s) + p(\vec{x}|b)}.
\end{equation}
The semi-positive-definiteness of $p(\vec{x}|s)$ and $p(\vec{x}|b)$ restricts $D(\vec{x})$ to the interval $[0,1]$ for all $\vec{x}$.  
\item The {\it sensitivity} ${\cal S}$ of a particular threshold $D_{\rm cut}$ on the discriminant function is defined as the reciprocal of $\overline{\sigma^{95\%}}$.  $D_{\rm cut}$ is chosen to maximize ${\cal S}$.  
\item The region of variable space having $D(\vec{x})>D_{\rm cut}$ is used to determine the actual 95\% CL cross section upper limit $\sigma^{95\%}$~\cite{sigma95}.
\end{list}
When provided with a signal model and a choice of variables $\cal V$, \Quaero\ uses this algorithm and D\O\ Run I data to compute an upper limit on the cross section of the signal.  Instructions for use are available from the \Quaero\ web site.

\section{Data}

Table~\ref{tbl:Data} shows the data available within \Quaero, and Table~\ref{tbl:Bkg} summarizes the backgrounds.  These data and their backgrounds are described in more detail in Ref.~\cite{SherlockPRD2}.  The final states are inclusive, with many events containing one or more additional jets.  Kolmogorov-Smirnov tests have been used to demonstrate agreement between data and the expected backgrounds in many distributions.  The fraction of events with true final state objects satisfying the cuts shown that satisfy these cuts after reconstruction is given as an ``identification'' efficiency ($\epsilon_{\rm ID}$).  Because electrons are more accurately measured and more efficiently identified than muons in the D\O\ detector, the corresponding muon channels $\mu\met\jj$ and $\mu\mu\jj$ have been excluded from these data.

\begin{table}[htb]
\centering
\begin{tabular}{lcrc}
Final State	& Selection criteria	& $\epsilon_{\rm ID}$ & $\int{{\cal L}\,dt}$ \\ \hline
$e\mu$ &
\parbox[t]{2.8cm}{$p_T^{e,\mu} > 15$~GeV \\ $\abs{\eta_{\rm det}^\mu}<1.7$} &
0.30 & $108\pm5$~pb$^{-1}$ \\ \hline
$e\met\jj$ &
\parbox[t]{2.8cm}{ $p_T^{e,j_{1,2}} > 20$~GeV \\ $\met > 30$~GeV \\ $p_T^{e\met}>40$~GeV } &
0.61 & $115\pm6$~pb$^{-1}$ \\ \hline
$ee\jj$ &
\parbox[t]{2.8cm}{ $p_T^{e_{1,2},j_{1,2}} > 20$~GeV } &
0.70 & $123\pm7$~pb$^{-1}$ \\ 
\end{tabular}
\caption{A summary of the data available within \Quaero, including the selection cuts applied and the efficiency of identification requirements.  The final states are inclusive, with many events containing one or more additional jets.  Reconstructed jets satisfy $p_T^{j}>15$~GeV and $\abs{\eta_{\rm det}^{j}}<2.5$, and reconstructed electrons satisfy $p_T^{e}>15$~GeV and ($\abs{\eta_{\rm det}^e}<1.1$ or $1.5<\abs{\eta_{\rm det}^e}<2.5$), where $\eta_{\rm det}$ is the pseudorapidity measured from the center of the detector.}
\label{tbl:Data}
\end{table}

\begin{table}[htb]
\centering
\begin{tabular}{l|ccccc}
  & \multicolumn{5}{c}{Standard model backgrounds} \\
Final State	& multijets & $W$ & $Z$ & $VV$ & $t\bar{t}$ \\ \hline
$e\mu$ & data & data & {\sc isajet} & {\sc pythia} & {\sc herwig} \\ \hline
$e\met\jj $ & data	& {\sc vecbos} & -- & {\sc pythia} & {\sc herwig} \\ \hline
$ee\jj$ 	& data & -- & {\sc pythia} & {\sc pythia} & -- \\
\end{tabular}
\caption{Standard model backgrounds (often produced with accompanying jets) to the final states considered.   $VV$ denotes $WW$, $WZ$, and $ZZ$; ``data'' indicates backgrounds from jets misidentified as electrons estimated using data.  Monte Carlo programs ({\sc isajet}~\protect\cite{isajet}, {\sc pythia}~\protect\cite{pythia}, {\sc herwig}~\protect\cite{herwig}, and {\sc vecbos}~\protect\cite{vecbos}) are used to estimate several sources of background.}
\label{tbl:Bkg}
\end{table}

To check standard model results, we remove $WW$ and $ZZ$ production from the background estimate and search (i) for standard model $WW$ production in the space defined by the transverse momentum of the electron ($p_T^e$) and missing transverse energy ($\met$) in the final state $e\mu\met$, and (ii) for standard model $ZZ$ production in the space defined by the invariant mass of the two electrons ($m_{ee}$) and two jets ($m_{jj}$) in the final state $ee\jj$.  Removing $t\bar{t}$ production from the background estimate, we search for this process (iii) in the final state $e\met \jjjj$ using the two variables laboratory aplanarity ($A$) and $\sum{p_T^j}$, and (iv) in the final state $e\mu\met\jj$, using the two variables $p_T^e$ and $\sum{p_T^j}$, assuming a top quark mass of 175~GeV.  

Including all standard model processes in the background estimate, we look for evidence of new heavy resonances.  We search (v) for resonant $WW$ production in the final state $e\met\jj$, using the single variable $m_{e\nu jj}$ after constraining $m_{e\nu}$ and $m_{jj}$ to $M_W$, and (vi) for resonant $ZZ$ production in the final state $ee \jj$, using the variable $m_{eejj}$ after constraining $m_{jj}$ to $M_Z$.  In both cases we remove events that cannot be so constrained.  To obtain a specific signal prediction, we assume that the resonance behaves like a standard model Higgs boson in its couplings to the $W$ and $Z$ bosons.  Constraining $m_{e\nu}$ to $M_W$ and $m_{jj}$ to $M_Z$, we use the quality of the fit and $m_{e\nu jj}$ to search (vii) for a massive $W'$ boson in the extended gauge model of Ref.~\cite{Altarelli}.  Using $m_{e\nu\jjjj}$ after constraining $m_{e\nu}$ to $M_W$, we search (viii) for a massive narrow $Z'$ resonance with $Z$-like couplings decaying to $t\bar{t} \rightarrow W^+bW^-\bar{b} \rightarrow e\nu\jjjj$. 

Non-resonant new phenomena are also considered.  The variables $m_{jj}$ and either $m^T_{e\nu}$ or $m_{ee}$ are used to search for a light Higgs boson produced (ix) in association with a $W$ boson, and (x) in association with a $Z$ boson.  Finally, we search (xi) for first generation scalar leptoquarks with mass 225~GeV in the final state $ee \jj$ using $m_{ee}$ and $S_T$, the summed scalar transverse momentum of all electrons and jets in the event.  The numerical results of these searches are listed in Table~\ref{tbl:examples1}.  Figures~\ref{tbl:examples2} and~\ref{tbl:examples3} present plots of the signal density, background density, and selected region in the variables considered.  

\begin{table}[htb]
\centering
\begin{tabular}{clcrr}
Process & \multicolumn{1}{c}{$\epsilon_{\rm sig}$} & $\hat{b}$ & \multicolumn{1}{c}{$N_{\rm data}$} & \multicolumn{1}{c}{$\sigma^{95\%}\times {\cal B}$} \\ \hline
$WW\rightarrow e\mu\met$	& \wwemusigeff	& $\wwemubkg\pm\wwemubkgerror$	& \wwemudata	& \wwemuxseclimit  \\
$ZZ\rightarrow ee\jj$		& \zzcsigeff	& $\zzcbkg\pm\zzcbkgerror$	& \zzcdata	& \zzcxseclimit \\ 
$t\bar{t}\rightarrow e\met\jjjj$	& \ttemetsigeff	& $\ttemetbkg\pm\ttemetbkgerror$	& \ttemetdata	& \ttemetxseclimit \\
$t\bar{t}\rightarrow e\mu\met\jj$	& \ttemusigeff	& $\ttemubkg\pm\ttemubkgerror$	& \ttemudata	& \ttemuxseclimit \\ \hline
$h_{175}\rightarrow WW\rightarrow e\met\jj$	& \hwwcasigeff	& $\hwwcabkg\pm\hwwcabkgerror$	& \hwwcadata	& \hwwcaxseclimit \\
$h_{200}\rightarrow WW\rightarrow e\met\jj$	& \hwwcbsigeff	& $\hwwcbbkg\pm\hwwcbbkgerror$	& \hwwcbdata	& \hwwcbxseclimit \\
$h_{225}\rightarrow WW\rightarrow e\met\jj$	& \hwwccsigeff	& $\hwwccbkg\pm\hwwccbkgerror$	& \hwwccdata	& \hwwccxseclimit \\
$h_{200}\rightarrow ZZ\rightarrow ee\jj$	& \hzzcasigeff	& $\hzzcabkg\pm\hzzcabkgerror$	& \hzzcadata	& \hzzcaxseclimit \\
$h_{225}\rightarrow ZZ\rightarrow ee\jj$	& \hzzcbsigeff	& $\hzzcbbkg\pm\hzzcbbkgerror$	& \hzzcbdata	& \hzzcbxseclimit \\
$h_{250}\rightarrow ZZ\rightarrow ee\jj$	& \hzzccsigeff	& $\hzzccbkg\pm\hzzccbkgerror$	& \hzzccdata	& \hzzccxseclimit \\ 
$W_{200}'\rightarrow WZ\rightarrow e\met\jj$ & \wprimeasigeff & $\wprimeabkg\pm\wprimeabkgerror$ & \wprimeadata & \wprimeaxseclimit \\ 
$W_{350}'\rightarrow WZ\rightarrow e\met\jj$ & \wprimedsigeff & $\wprimedbkg\pm\wprimedbkgerror$ & \wprimeddata & \wprimedxseclimit \\ 
$W_{500}'\rightarrow WZ\rightarrow e\met\jj$ & \wprimegsigeff & $\wprimegbkg\pm\wprimegbkgerror$ & \wprimegdata & \wprimegxseclimit \\ 
$Z_{350}'\rightarrow t\bar{t}\rightarrow e\met\jjjj$	& \zttasigeff	& $\zttabkg\pm\zttabkgerror$	& \zttadata	& \zttaxseclimit \\ 
$Z_{450}'\rightarrow t\bar{t}\rightarrow e\met\jjjj$	& \zttesigeff	& $\zttebkg\pm\zttebkgerror$	& \zttedata	& \zttexseclimit \\ 
$Z_{550}'\rightarrow t\bar{t}\rightarrow e\met\jjjj$	& \zttgsigeff	& $\zttgbkg\pm\zttgbkgerror$	& \zttgdata	& \zttgxseclimit \\ \hline
$Wh_{115}\rightarrow e\met\jj$	& \whsigeff	& $\whbkg\pm\whbkgerror$	& \whdata	& \whxseclimit \\
$Zh_{115}\rightarrow ee\jj$	& \zhsigeff	& $\zhbkg\pm\zhbkgerror$	& \zhdata	& \zhxseclimit \\ 
$LQ_{225}\overline{LQ}_{225}\rightarrow ee\jj$	& \lqcsigeff	& $\lqcbkg\pm\lqcbkgerror$	& \lqcdata	& \lqcxseclimit \\
\end{tabular}
\caption{Limits on cross section $\times$ branching fraction for the processes discussed in the text.  All final states are inclusive in the number of additional jets.  The fraction of the signal sample satisfying \Quaero's selection criteria is denoted $\epsilon_{\rm sig}$; $\hat{b}$ is the number of expected background events satisfying these criteria; and $N_{\rm data}$ is the number of events in the data satisfying these criteria.  The subscripts on $h$, $W'$, $Z'$, and $LQ$ denote assumed masses, in units of GeV.}
\label{tbl:examples1}
\end{table}

\newcommand{\OnOffepsfbox}[1]
{\epsfxsize=3.5in\raisebox{-0.3in}[0.9in][0.3in]{\epsfbox{#1}}}
\begin{figure}[ht]
\centering
\begin{tabular}{ccc}
Background density & \ \ \ \ Signal density & Selected region \\ 
(a) & \ \ (b) & (c) \\ 
\multicolumn{3}{c}{\OnOffepsfbox{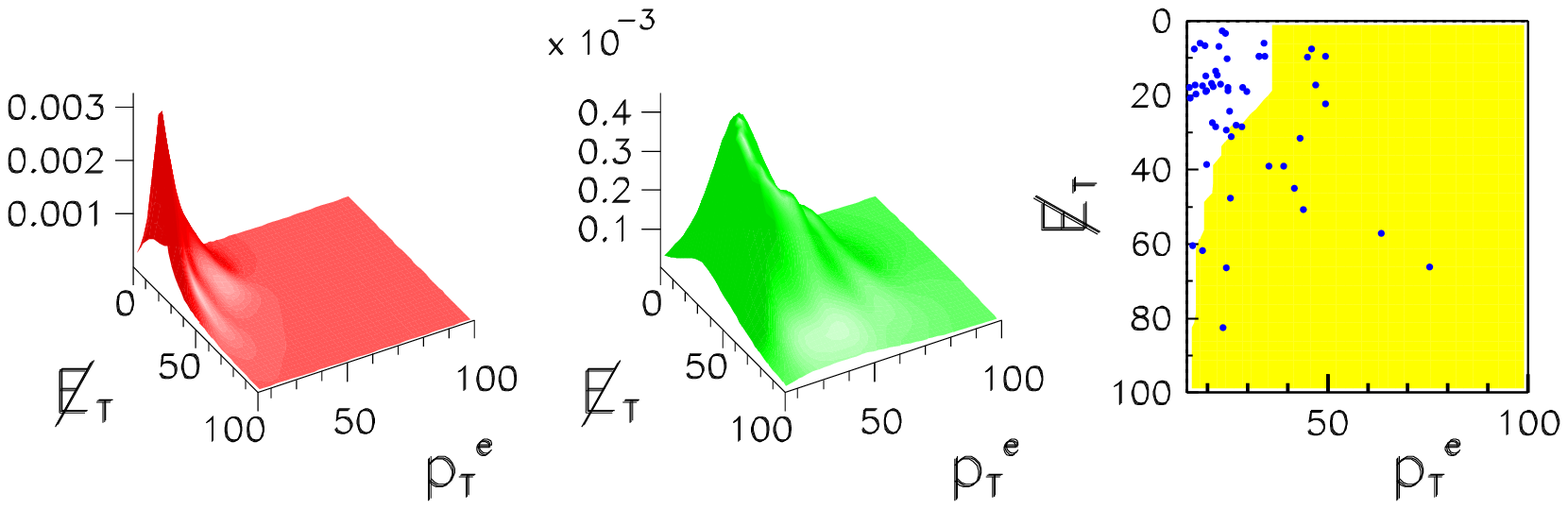}} \\
\multicolumn{3}{c}{\OnOffepsfbox{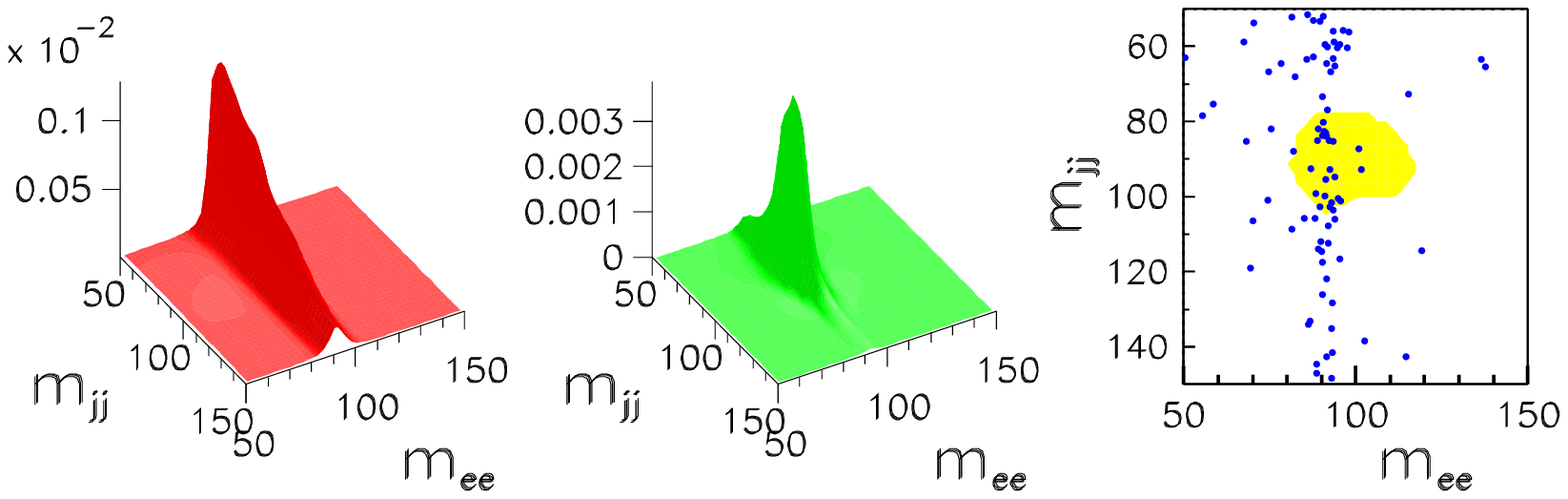}} \\
\multicolumn{3}{c}{\OnOffepsfbox{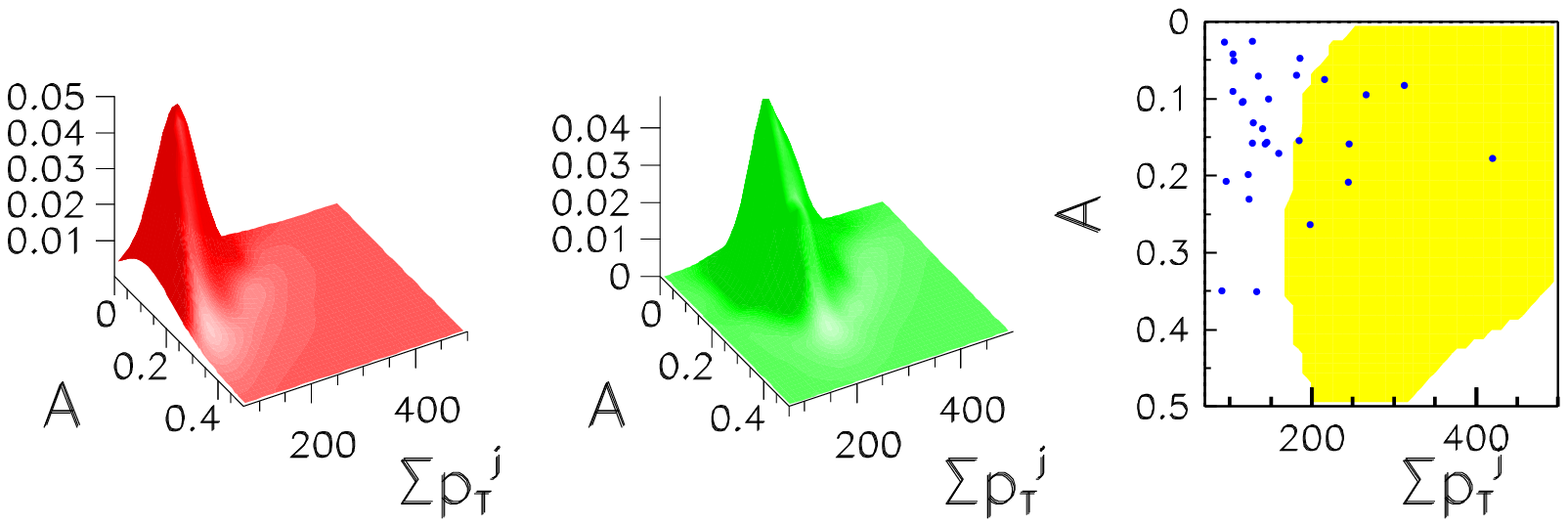}} \\
\multicolumn{3}{c}{\OnOffepsfbox{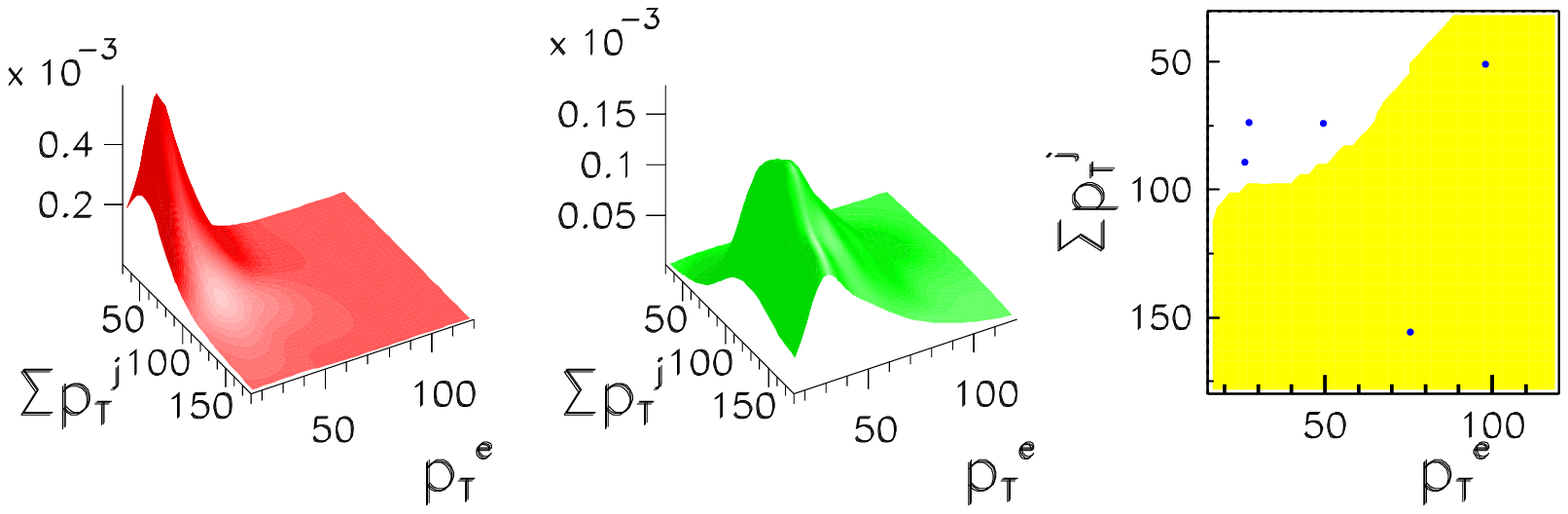}} \\
\end{tabular}
\caption{The background density (a), signal density (b), and selected region (shaded) (c) determined by \Quaero\ for the standard model processes discussed in the text.   From top to bottom the signals are: $WW\rightarrow e\mu\met$, $ZZ\rightarrow ee\jj$, $t\bar{t}\rightarrow e\met\jjjj$, and $t\bar{t}\rightarrow e\mu\met\jj$.  The dots in the plots in the rightmost column represent events observed in the data.}
\label{tbl:examples2}
\end{figure}

\begin{figure}[ht]
\centering
\begin{tabular}{ccc}
Background density & \ \ \ \ Signal density & Selected region \\ 
(a) & \ \ (b) & (c) \\ 
\multicolumn{3}{c}{\OnOffepsfbox{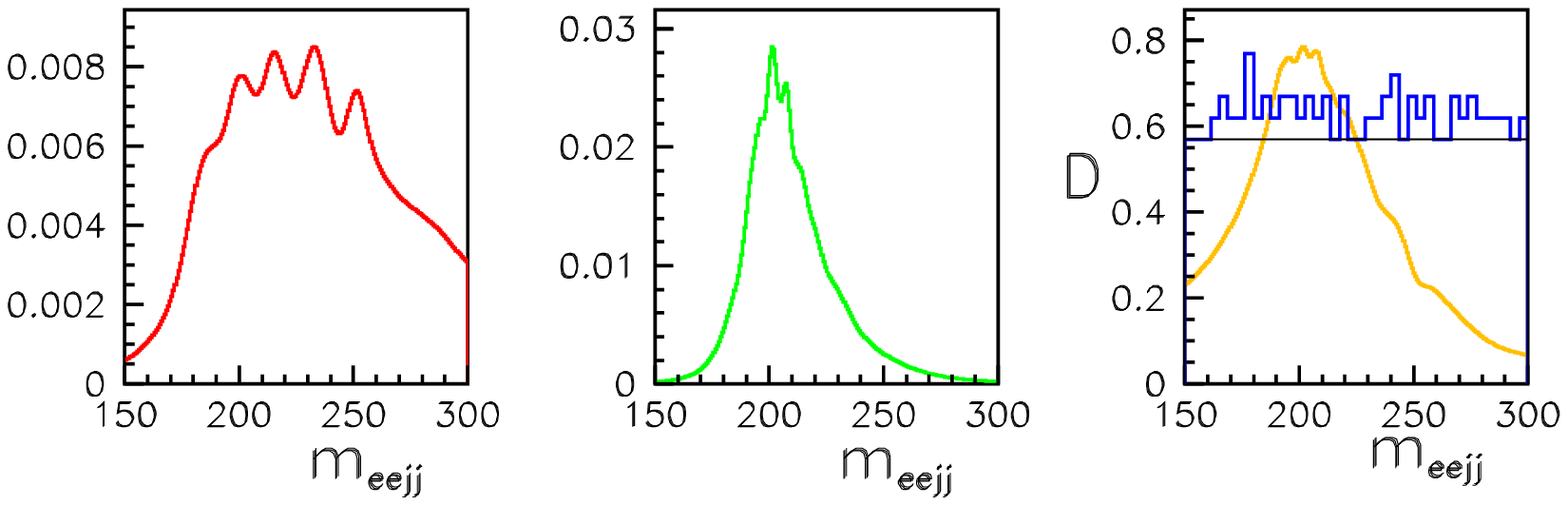}} \\
\multicolumn{3}{c}{\OnOffepsfbox{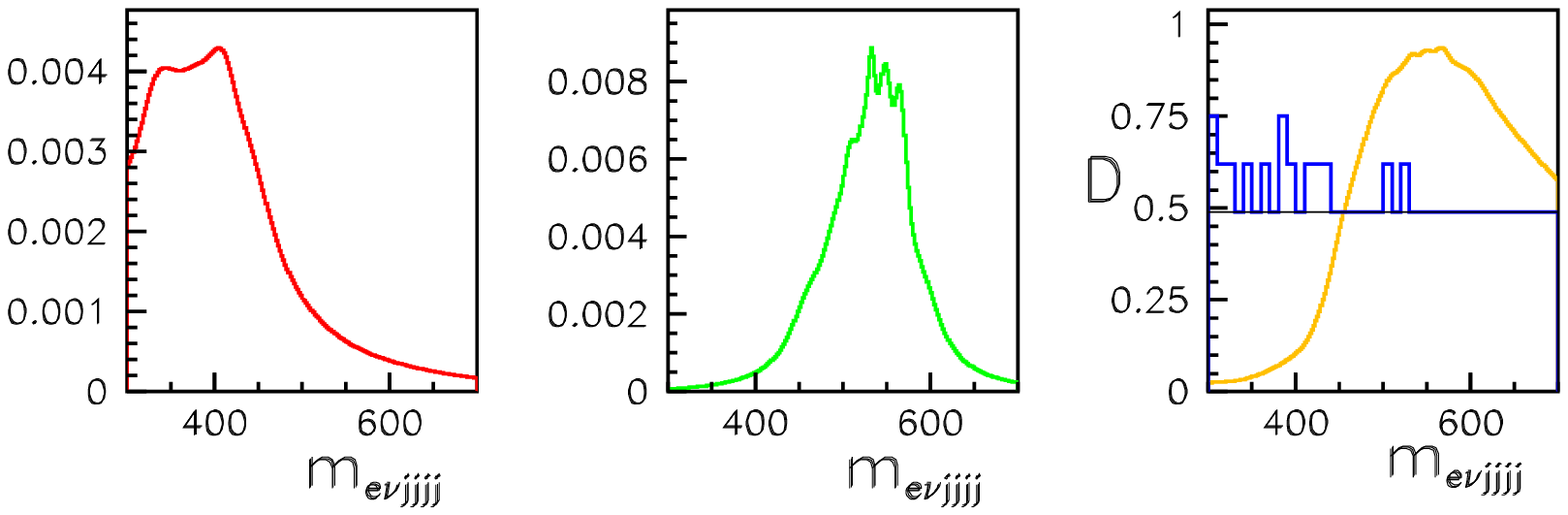}} \\
\multicolumn{3}{c}{\OnOffepsfbox{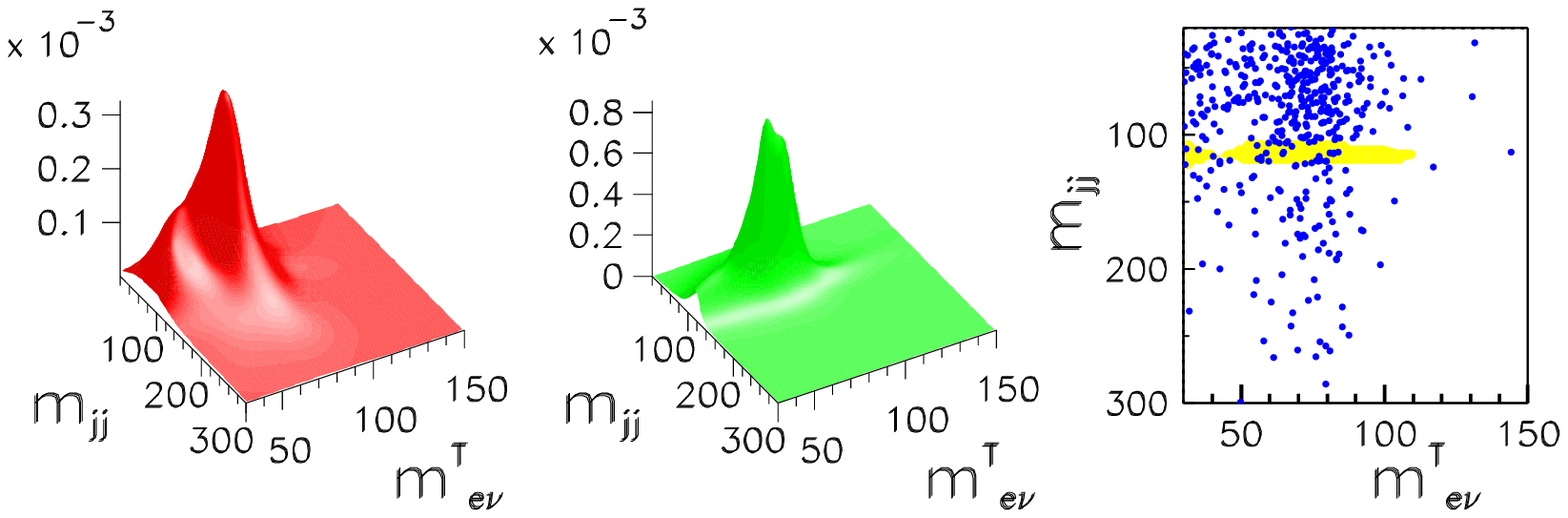}} \\
\multicolumn{3}{c}{\OnOffepsfbox{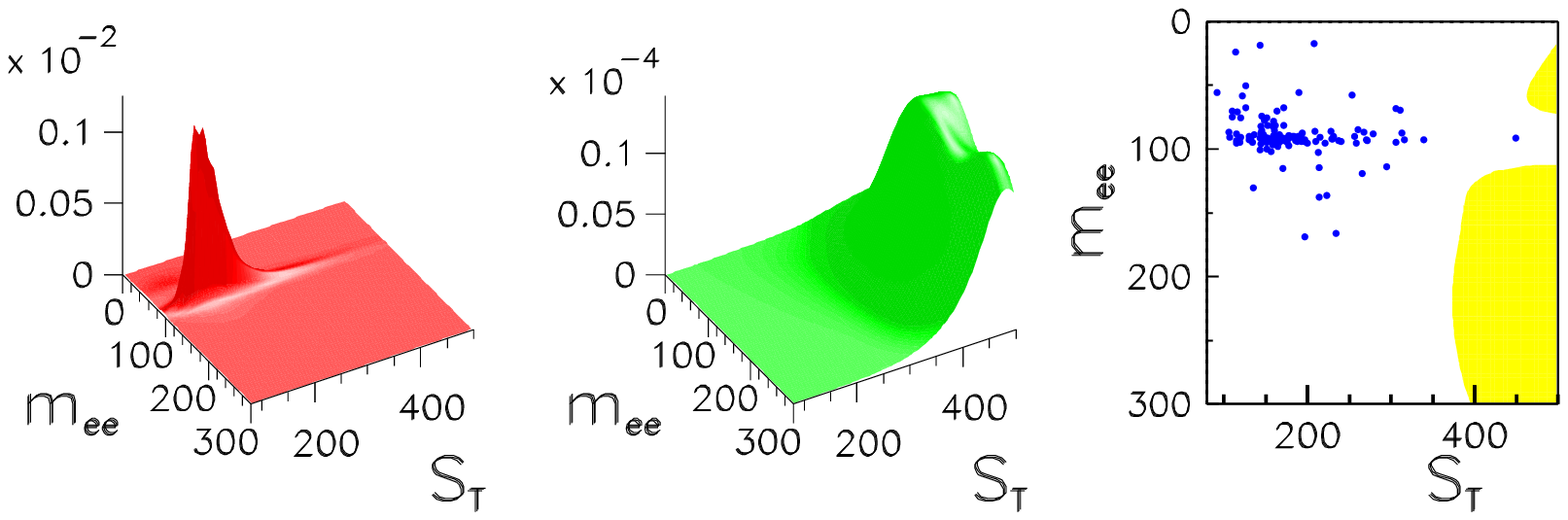}} \\
\end{tabular}
\caption{\Quaero's analysis of signatures involving undiscovered particles.  From top to bottom the hypothetical signals are: $h_{200}\rightarrow ZZ \rightarrow ee\jj$, $Z'_{550}\rightarrow t\bar{t} \rightarrow e\met\jjjj$, $Wh_{115} \rightarrow e\met\jj$, and $LQ_{225}\overline{LQ}_{225} \rightarrow ee\jj$.  Plots (c) of the first two rows show the discriminant $D$ (curve), the threshold $D_{\rm cut}$ (horizontal line), and the data (histogram); the region with $D>D_{\rm cut}$ is selected.}
\label{tbl:examples3}
\end{figure}

We note slight indications of excess in the searches for $t\bar{t}\rightarrow e\met\jjjj$ and $t\bar{t}\rightarrow e\mu\met\jj$ (corresponding to cross section $\times$ branching fractions of $\sigma\times {\cal B} = \ttemetxsec^{+\ttemetxsecUpperError}_{-\ttemetxsecLowerError}$~pb and $\ttemuxsec^{+\ttemuxsecUpperError}_{-\ttemuxsecLowerError}$~pb) that are consistent with our measured $t\bar{t}$ production cross section of $5.5\pm1.8$~pb~\cite{topCrossSection} and known $W$ boson branching fractions.  Observing no compelling excess in any of these processes, limits on $\sigma\times {\cal B}$ are determined at the 95\% CL.  As expected, we find these data insensitive to standard model $ZZ$ production (with predicted $\sigma\times {\cal B} \approx 0.05$~pb), and to associated Higgs boson production (with predicted $\sigma\times {\cal B} \ltapprox 0.01$~pb).  As a check of the method, \Quaero\ almost exactly duplicates a previous search for $LQ\overline{LQ} \rightarrow ee\jj$~\cite{LeptoquarksToEE}.

\Quaero\ is a method both for automatically optimizing searches for new physics and for allowing D\O\ to make a subset of its data available for general use.  In this Letter we have outlined the algorithm used in \Quaero, and we have described the final states currently available for analysis using this method. \Quaero's performance on several examples, including both standard model and resonant $WW$, $ZZ$, and $t\bar{t}$ production, has been demonstrated.  The limits obtained are comparable to those from previous searches at hadron colliders, and the search for $W'\rightarrow WZ$ is the first of its kind.  This tool should increase the facility with which new models may be tested in the future.

%
We thank the staffs at Fermilab and collaborating institutions, 
and acknowledge support from the 
Department of Energy and National Science Foundation (USA),  
Commissariat  \` a L'Energie Atomique and 
CNRS/Institut National de Physique Nucl\'eaire et 
de Physique des Particules (France), 
Ministry for Science and Technology and Ministry for Atomic 
   Energy (Russia),
CAPES and CNPq (Brazil),
Departments of Atomic Energy and Science and Education (India),
Colciencias (Colombia),
CONACyT (Mexico),
Ministry of Education and KOSEF (Korea),
CONICET and UBACyT (Argentina),
The Foundation for Fundamental Research on Matter (The Netherlands),
PPARC (United Kingdom),
Ministry of Education (Czech Republic),
and the A.P.~Sloan Foundation.

\bibliographystyle{unsrt}

\begin{thebibliography}{2}
\bibitem{sigma95}
$\sigma^{95\%}$ is a Bayesian limit, computed assuming a flat prior.
\end{thebibliography}

\end{document}